# BESTOpt: A Modular, Physics-Informed Machine Learning based Building Modeling, Control and Optimization Framework

Zixin Jiang[1], Ruizhi Song[2], Guowen Li[2], Yuhang Zhang[2], Zheng O'Neill[2], Xuezheng Wang[1], Judah Goldfeder[3], Bing Dong[1]

1 Department of Mechanical and Aerospace Engineering, Syracuse University, Syracuse, NY 13210, USA

2 J. Mike Walker '66 Department of Mechanical Engineering, Texas A&M University, College Station, TX 77843, USA

3 Department of Mechanical Engineering, Columbia University, New York, NY 10027, USA

**Abstract**

Modern buildings are increasingly interconnected with occupancy, heating, ventilation, and air-conditioning (HVAC) systems, distributed energy resources (DERs), and power grids. Modeling, control, and optimization of such multi-domain systems play a critical role in achieving building-sector decarbonization. However, most existing tools lack scalability and physical consistency for addressing these complex, multi-scale ecosystem problems. To bridge this gap, this study presents BESTOpt, a modular, physics-informed machine learning (PIML) framework that unifies building applications, including benchmarking, evaluation, diagnostics, control, optimization, and performance simulation. The framework adopts a cluster–domain–system/building–component hierarchy and a standardized state–action–disturbance–observation data typology. By embedding physics priors into data-driven modules, BESTOpt improves model accuracy and physical consistency under unseen conditions. Case studies on single-building and cluster scenarios demonstrate its capability for multi-level centralized and decentralized control. Looking ahead, BESTOpt lays the foundation for an open, extensible platform that accelerates interdisciplinary research toward smart, resilient, and decarbonized building ecosystems.

## 1. Introduction

### 1.1 Background and Motivation

The building sector plays an increasingly vital role in the global transition toward decarbonization. To meet the 2050 targets, approximately 10,000 buildings per day will need to be decarbonized over the next 25 years in the United States (Wetter, M., & Sulzer, M. 2024). This urgent demand calls for a paradigm shift toward a holistic framework that unifies modeling, control, and optimization.

On one hand, such a framework is essential for helping stakeholders, such as building owners, facility managers, architects, and policymakers to better understand building performance and how to design, construct, retrofit, maintain, diagnose, and operate buildings optimally (Jiang, Z., & Dong, B. 2024).

On the other hand, future buildings are evolving into highly interconnected ecosystems that combine occupants, buildings, HVAC systems, DERs, transportation, the electric grid, and urban infrastructure. These next-generation buildings must not only be energy-efficient but also capable of sensing, learning, and adapting in real time to balance energy efficiency, resilience, flexibility, and occupant well-being under dynamic conditions.

This shift motivates us to develop a unified, holistic framework that can simulate, model, and optimize interconnected building energy systems that include thermal dynamics, occupancy behaviors, HVAC operations, DER dispatch, and grid interactions—in a physics-aware, scalable and generalizable manner for multiscale building energy applications across benchmarking and evaluation, sensing and diagnostics, controls and optimization, building performance simulation, and measurement and verification.

### 1.2 Research Gaps

Despite significant advancements in sensing technologies, building automation, data-driven methods, and control strategies, current approaches for building modeling, control, and optimization do not yet support the realization of the above-mentioned holistic framework. Here, we identify several critical research gaps:

- **Tradeoff Between Scalability and Physics Consistency**

Existing approaches either rely heavily on physics-based models like EnergyPlus and Modelica, which offer high physical fidelity but are often complicated, require detailed metadata, manual configuration, and it is challenging to calibrate these physics-based models, making them impractical for large-scale deployment. In contrast, data-driven methods such as machine learning and deep learning models are scalable and adaptive but often lack physics consistency and generalization ability under unseen conditions, especially when training data is limited, noisy, and sparse. This limits their reliability for benchmarking across diverse scenarios or transferring control strategies to new buildings.

- **Lack of Coordination and Interaction Modeling**

Existing platforms for building applications, such as CityLearn, REopt, HOMER Pro, and DER-CAM, optimize DER operation using pre-defined or predicted building load profiles, instead of detailed building/HVAC dynamic models. These tools typically treat buildings as static nodes instead of dynamic systems, neglecting real-time interactions between indoor conditions and control systems. For example, instead of using feedback control based on measured zone temperature (e.g., adjusting fan speed or valve positions), they assume fixed energy demands to be met. This simplification could lead to unrealistic system behavior and performance deviations. Furthermore, most existing frameworks focus on small-scale or decentralized control, treating each building as an isolated unit. This limits their ability to model and optimize interconnected urban systems where building-level and cross-domain interactions, and potentially centralized coordination are increasingly critical.

- **Limited Transparency and Non-Differentiable Architectures**

Most existing platforms are designed as opaque, non-transparent systems that offer limited access to internal components. This restricts adaptability, modularity making it difficult for users to inspect, modify, or extend individual components such as HVAC, DERs, or controllers. The absence of differentiable architectures further limits the integration of advanced, gradient-based optimization approaches. This challenge is particularly relevant for model-based control methods such as Model Predictive Control (MPC), which require explicit and differentiable system representations. Although model-free control (e.g., reinforcement learning) does not require differentiability, the lack of transparent and modular component models still poses practical challenges. Without clear access to component-level dynamics (e.g., heat pump, fan dynamics, thermal storage behavior), it becomes difficult to construct meaningful state representations, accelerate training, interpret learned policies, or ensure safe operation. As a result, both model-based and model-free approaches are constrained by the opacity of existing platforms.

**1.3 Our Contribution**

To address these challenges, we propose BESTOpt, **a modular, PIML based runtime environment for modeling, control, and optimization of interconnected occupancy-building–HVAC–DER–grid systems**. This framework enables scalable, physically consistent simulation and control of interconnected building energy systems through the following core innovations:

- **PIML based Module Development for Generalization and Consistency**

PIML represents a state-of-the-art modeling approach that integrates fundamental physical principles such as conservation laws, symmetries, and causal relationships with advanced machine learning techniques. This integration occurs through modifications to model architectures, loss functions, model parameters, or training algorithms. PIML combines the advantages of both physics-based models and data-driven models, demonstrating superior capabilities in enhancing generalization to unseen scenarios, ensuring physical consistency, and reducing data requirements. By leveraging these techniques, we develop PIML-based modules as the fundamental core stones of our framework. These modules can learn from complex, interconnected building energy ecosystems while preserving physical consistency and interpretability.

- **Hierarchical Architecture for Cross-Domain and System-Level Interaction**

Our framework builds upon a hierarchical structure organized as cluster–domain–system/building–component levels, categorized by model states, actions, disturbances, and observations across multiple scales. This architecture facilitates modular component registration, efficient data management, and seamless interactions across different domains (e.g., thermal, electrical) and systems (e.g., HVAC, DER, occupants). This design enables comprehensive modeling of cross-domain interactions (e.g., integrating HVAC and building thermal dynamics into electrical optimization loops) and multi-scale coordination (e.g., from decentralized individual control to centralized coordinated control to hybrid approaches) that are critical for HVAC-building-DER-grid co-optimization.

- **Modular Design Enabling Flexibility, Transparency, and Differentiability**

The modular structure of our framework improved development flexibility by supporting independent module updates, sharing, replacement, and extension. This modularity enables scalable deployment across diverse buildings and applications while maintaining transparency and interpretability. Furthermore, each module is implemented in a differentiable form, enabling seamless integration with gradient-based optimization techniques, including MPC, reinforcement learning, and differentiable predictive control. This differentiability addresses a critical limitation of existing platforms and opens new possibilities for advanced control strategies.

## 2. Methodology

As shown in Figure 1, the BESTOpt framework is organized as a modular runtime environment consisting of four major parts:
(1) Disturbance Module, which generates time-varying external inputs for each simulation step;
(2) PIML Dynamic Module Library, which includes interconnected building, HVAC, and DER modules;
(3) Controller Module, which manages the operation of these dynamic systems through control logic; and
(4) Observation Module, which collects and organizes system data during runtime.

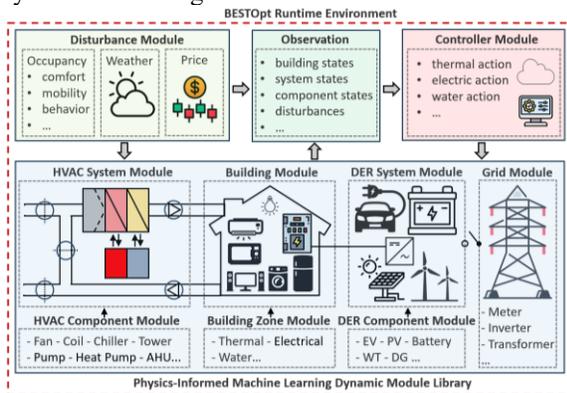

Figure 1 Overall Structure of BESTOpt

### 2.1 Module Hierarchy and Data Typology

Before diving into technical details, let us look ahead: Modern building energy systems are evolving into highly interconnected ecosystems that must simultaneously address **multi-scale** and **multi-domain** challenges:

- **Multi-scale Support**

The framework must support modeling, control, and optimization across multiple levels of granularity from individual components (e.g., fans, coils) to system-level entities (e.g., HVAC, DERs), and up to building clusters comprising multiple buildings and interconnected systems.

- **Multi-domain Integration**

The environment must support multi-domain integrations. This includes decentralized simple cases such as a single HVAC system serving one building or a standalone DER system, as well as more complex setups where, for example, a centralized HVAC system serves multiple buildings, or a DER system supplies power to a group of buildings. Moreover, it should support cross-domain coordination, such as joint control of HVAC and DER systems to enable holistic energy management.

With these requirements in mind, we explain the concept and structure of BESTOpt framework in the following.

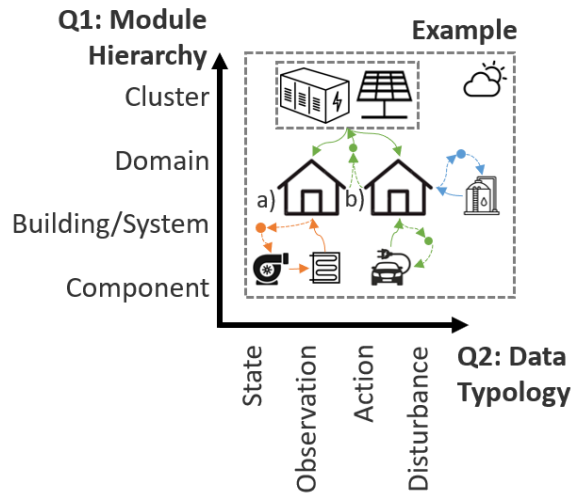

Figure 2 Configuration example of a cluster. We first introduce the module hierarchy and data typology, and then present a simple example to show how the proposed structure operates in practice

#### 2.1.1 Hierarchical Modular Structure

To support such complexity, the BESTOpt framework adopts a **cluster–domain–system/building–component** hierarchy to manage coordination across different scales and domains as shown in Figure 2 (Y axis):

- **Cluster level**

This is the topmost level in BESTOpt, representing the overall simulation scope (e.g., a residential cluster, campus, or city). It coordinates interactions among multiple domains, manages shared resources, and executes global optimization or coordination strategies. Each cluster has a unique cluster ID, and a simulation environment may contain multiple clusters.

- **Domain level**

This level defines the type of energy flow within a cluster. The current version supports three energy domains: thermal, electrical, and water. The framework can be extended to support others, such as natural gas or hydrogen in the future. Regardless of the number of systems or buildings, all energy flows are categorized under one of these three domains.

- **System/Building level**

To maintain a clean environment structure and ensure that systems can be easily shared, updated, and replaced with minimal effort, the next level in the hierarchy comprises systems (energy suppliers) and buildings (energy demanders). This level represents well-packaged entities composed of multiple components, with each system assigned a unique system ID. Notably, a single system can influence multiple domains; for example, an HVAC system may consume electricity while producing heated air and domestic hot water. Similar, buildings typically link to all three domains, with thermal, electrical, and water demands.

- **Component level**

This level represents the fundamental unit used to construct systems. For instance, a fan-coil HVAC system may include fans, coils, pumps, chillers, and towers, depending on the configuration setup. Similarly, a building consists of thermal, water, and electrical components for demand calculation. DER systems may include multiple EVs, batteries, and PVs, with each component assigned a unique component ID.

This hierarchical modular structure forms the foundation of the BESTOpt environment, enabling consistent configuration and management, cross-domain data exchange and coordination and flexible scaling from single components to multi-building/system clusters.

*2.1.2 Data Typology and Flow*

To support modular interoperability and ensure consistent data exchange across layers, BESTOpt standardizes all information through a unified **state–action–disturbance–observation** data typology as shown in Figure 2 (X axis). This framework defines how data is collected, stored, and managed across all modules, as well as how it flows within the hierarchical structure.

*2.1.2.1 Data Typology*

BESTOpt organizes all module-level information into the following four categories:

- **State**

Represents internal, time-varying component/system conditions that define dynamic behaviors.
*Examples:* zone air temperature, battery state-of-charge (SOC), chiller water temperature.

- **Action**

Refers to control or decision variables sent by a controller to execute components/systems.
*Examples:* supply air temperature setpoint, supply air flowrate setpoint, pump water flowrate setpoint.

- **Disturbance**

Denotes external or exogenous inputs that impact the system but are not directly controlled. These include environmental, operational, and economic factors.
*Examples:* outdoor air temperature, solar radiation, occupancy status, electricity prices.

- **Observation**

Includes all data collected via sensors and meters, inferred from models, predicted by forecasting models or aggregated from simulation outputs. Observations are used for control decisions, and performance monitoring.
*Examples:* measured indoor temperature, metered power consumption, predicted occupancy status.

This typology enables consistent treatment of heterogeneous data across different modules and supports hybrid modeling paradigms.

*2.1.2.2 Data Flow*

The flow of data in BESTOpt follows the hierarchical modular structure described earlier. Each of the data types such as state, action, and observation can exist at multiple levels within the cluster–domain–system/building–component hierarchy.

At runtime, data at higher levels are aggregated from lower levels through pre-defined functions such as aggregation. Conversely, high-level control actions can be decomposed and dispatched downward to individual systems or components. For example, a building's total electric load is computed by aggregating the power consumption of its subcomponents, such as appliances, lighting, and the parallel HVAC systems. Similarly, a control signal at the cluster level (e.g., power flow among DERs) can be disaggregated into system-level setpoints and further propagated to component-level actuators.

Regarding disturbance, BESTOpt currently assumes a shared, global disturbance signal across all modules within the environment. For example, a single weather disturbance. However, the framework allows for localized disturbances (e.g., using different weather files for different clusters) when available, especially at the urban scale when local climate matters. Despite this, the underlying disturbance handling mechanism remains centralized through a single disturbance module that can broadcast tailored signals to multiple domains or clusters.

This layered data typology and flow mechanism ensures consistent abstraction, traceable signal propagation, and modular compatibility across the BESTOpt runtime environment. It also lays the groundwork for advanced features such as reinforcement learning, centralized control optimization and fine-grained performance evaluation across multiple scales.

### 2.1.3 An Example to Configure a Building Cluster

Figure 2 illustrates an example cluster configuration containing two buildings that share a centralized PV–battery DER system. Building A is equipped with a fan coil HVAC unit, while Building B includes an EV and a water tank. This simplified example is used to demonstrate how the proposed hierarchical structure and data typology work within the framework.

At the component level, basic units such as fans and coils form an HVAC system, which is registered under the thermal domain (orange arrows). Similarly, the EV and water tank are components registered under the electrical and water domains (green and blue arrows), respectively. The shared PV–battery system forms a DER system registered in the electrical domain.

Each building contains thermal, electrical, and water subsystems that calculate corresponding demands and are registered under their respective domains. Together, these three domains constitute the entire cluster, allowing flexible management of all components within it.

Regarding data typology, this cluster shares a common disturbance, while the states of each component, system, and building are stored as state variables. Measurement and forecasted data are treated as observations, and control decisions are represented as actions. Each hierarchical level across from component, system, building, and cluster maintains its own set of state, action, and observation variables.

## 2.2 Module Description

The BESTOpt framework consists of three major categories of modules, each playing a distinct role in simulating and coordinating:

- **Dynamics Modules** – Representing basic physical systems such as HVAC, DERs and buildings, built upon on our open-source PIML module library.
- **Controller Modules** – Responsible for generating control actions at various levels by processing observations and upstream actions.
- **Disturbance Modules** – Used to generate exogenous inputs such as weather, occupancy, and pricing signals.

### 2.2.1 Introduction of the Base Module Class

We first introduce the shared base class of BESTOpt, from which all modules inherit, providing a consistent interface and defining three core functions essential for simulation execution:

- **Initialize ():** Prepares the module for simulation by instantiating its internal structure and setting initial states.
- **Reset ():** Resets the module to its initial condition, enabling reproducibility and repeated simulations.
- **Step ():** Executes one simulation time step to calculate the updated dynamic state outputs.

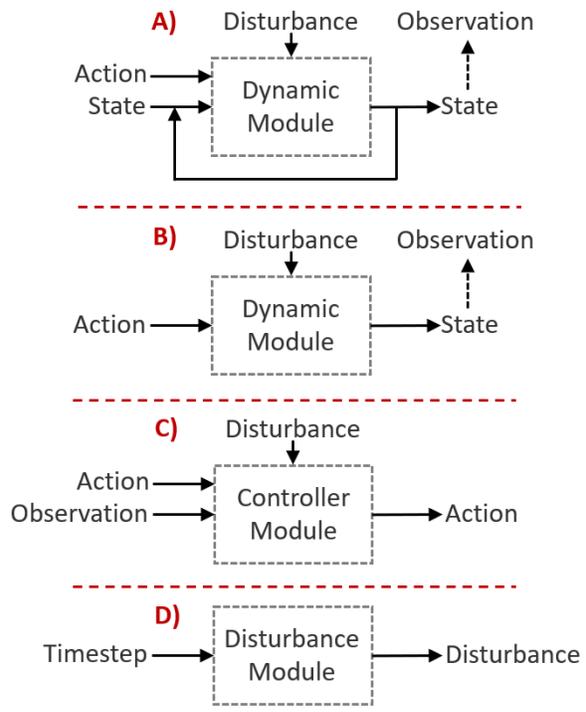

*Figure 3 Different step functions for dynamic, controller and disturbance modules*

Here, we describe the detailed step function for each module:

### 2.2.1.1 Step Function of Dynamic Modules (Figure 3 A and B)

There are two types of dynamic modules:

- **Stateful dynamic modules** (e.g., building thermal models, coils, and chillers) simulate processes where the system's next state is dependent on its current state. These modules maintain internal memory and evolve over time.
- **Stateless dynamic modules** (e.g., fans and pumps) assume no memory of previous states. Their outputs at each timestep are determined solely by the current inputs.

In both cases, dynamic modules accept disturbances and control actions as inputs and output updated state variables for the next simulation step.

### 2.2.1.2 Step Function of Controller Modules (Figure 3 C)

Controller modules generate control decisions by processing inputs such as local observations and upstream control signals. For example, a component-level controller uses a reference signal from a system-level controller to compute local actions, while a domain-level controller aggregates information from all systems within the domain to coordinate higher-level decisions.

### 2.2.1.3 Step Function of Disturbance Modules (Figure 3 D)

These modules produce time-varying exogenous inputs such as occupancy profiles, weather data, or economic signals. They operate based on the current simulation timestep and are shared globally across the environment, serving as inputs to all relevant modules.

A key architectural principle in BESTOpt is the hierarchical modular interaction, which enforces a structured, disciplined approach to data exchange: Except the globally shared disturbance module, the rest modules follow a directional input-output constraint:
- Inputs can only be received from modules at the same or higher hierarchical level.
- Outputs can only be sent to modules at the same or lower level.

This structure ensures clean boundaries between modules and promotes reusability and composability. For example:
- A component-level controller accepts actions from the system-level controller and uses them as reference signals to compute local control actions.
- A domain-level controller has access to all relevant system-level information within its domain (e.g., thermal or electrical).
- A cluster-level controller, located at the top of the hierarchy, can aggregate cross-domain observations and coordinate control strategies across multiple domains, including multiple buildings and systems.

Such a design helps avoid unnecessary memory usage, reduces data redundancy, and provides a systematic framework for managing control and information flow. Users can flexibly define the control hierarchy based on their specific application needs: whether focusing on low-level actuation or high-level, e.g., grid coordination.

Overall, this unified module interface and hierarchical constraint logic enable seamless orchestration within the BESTOpt runtime environment, supporting scalable simulations, modular controller design, and advanced optimization techniques.

## 2.2 Introduction of the Dynamic Modules

The development of dynamic module library is an ongoing project. Currently, it includes physics-informed models for building thermal dynamics, as well as high-fidelity, physics-based modules for HVAC systems and distributed energy resources (DERs). These modules are designed to operate at various spatial and temporal resolutions and can be flexibly assembled based on simulation or control needs.

### 2.2.2.1 Building Models

Each building can be modeled as a collection of three conceptual components in BESTOpt, referred to as the thermal network, electrical network, and water network. They serve as logical abstractions within the building model that represent the building's energy demands and dynamic behaviors across their respective domains.

- The electrical network and water network modules calculate time-varying demand profiles for plug loads, lighting, domestic hot water (DHW), etc. These demands are driven by occupant behavior, which comes from external disturbance inputs.
- The thermal network module interacts with the HVAC system in real-time, rather than relying on the above-mentioned existing tools that use pre-calculated heating and cooling load curves. The zone temperature is updated dynamically based on internal gains, weather conditions and HVAC supply side outputs. In this formulation, the building does not passively receive HVAC operation signals; instead, it actively closes the control loop with the HVAC system, adjusting its thermal state and generating new control signals for HVAC system as needed.

We employ our previously developed Physics-Informed Modularized Neural Network (PI-ModNN) (Jiang, Z., & Dong, B. 2024, 2025) to model the thermal dynamics of buildings, as illustrated in Figure 4. PI-ModNN decomposes the overall heat transfer process into separate neural network modules, each responsible for learning distinct physical components. Physics-informed constraints and loss functions are applied during training to ensure physical consistency and improve generalization. This modular design philosophy not only enables accurate and interpretable modeling of thermal zones but also serves as the foundational concept that inspired the extension of BESTOpt beyond individual buildings to a broader ecosystem, which integrates occupancy, building, HVAC, DER, and grid systems through interconnected modular neural networks.

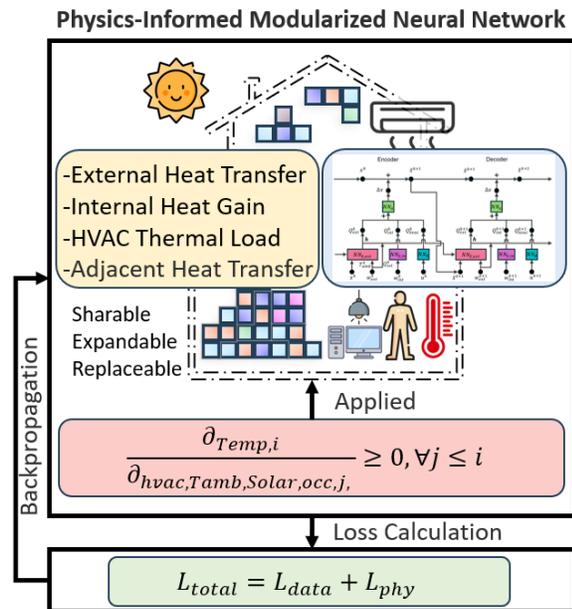

*Figure 4 Physics-Informed Modularized Neural Network for thermal dynamic modeling*

This modular approach not only enables accurate and interpretable modeling of thermal zones but also serves as the foundational concept that inspired us to extend the framework beyond buildings to the broader BESTOpt ecosystem.

In practice, PI-ModNN is formulated as a time-stepping model that predicts the zone temperature at the next time step based on the current state (e.g., zone air temperature), external disturbances (e.g., weather, occupancy), and control inputs (e.g., HVAC thermal load calculated using the supply air temperature and flow rate from the HVAC modules). The corresponding HVAC load is dynamically computed by the HVAC module, which interacts with the thermal model in real time. Further technical details such as model architecture, training procedures, and performance evaluation for both simulation and control are available in our prior work (Jiang, Z., & Dong, B. 2024, 2025).

*2.2.2.2 HVAC Systems*

The HVAC system is an aggregation of subcomponent-level HVAC modules, providing a unified representation of air and hydronic loop handling for building indoor environmental control. It receives setpoints from the system-level controller, such as the supply air temperature and airflow setpoints. The outputs of the HVAC module include the resulting supply air temperature and airflow rate, which are fed back to the thermal network model, as well as the system's energy consumption.

The module incorporates key physical components such as coils, fans, pumps, chillers, cooling towers, heat pumps, etc. Its modular architecture allows a flexible assembly and analysis of various HVAC configurations as additional components are developed and integrated. Furthermore, the framework supports modeling approaches that span from empirical component formulations to learning-based representations (Zhang et al., 2024). This flexibility provides a foundation for future integrations of PIML models, which combine data adaptability with physical consistency to enable more customized, generalizable, and robust prediction and control of HVAC systems.

- Fan: it drives air movement.
- Pump: it drives water circulation.
- Coil: it performs heat exchange between the water and air loops, currently represented as an effectiveness-based model.
- Chiller: it provides centralized cooling generation and currently supports either a Carnot COP-based model or a performance-curve-based model.
- Cooling tower: it provides condenser water cooling by rejecting heat from the condenser loop to the ambient air.
- Ice storage tank: it stores cooling energy in the form of latent heat by freezing and melting water, thereby enabling load shifting and peak demand reduction.
- Boiler: it heats water to provide space heating and meet domestic hot water demand.
- Heat pump: it transfers heat from the air, ground, or water to provide space heating and cooling.

*2.2.2.3 DER Systems*

DER systems in BESTOpt are classified into two primary types:
- Energy generation systems, such as photovoltaic (PV) panels, wind turbines, and diesel generators;

- Energy storage systems, including stationary battery storage and mobile electric vehicle (EV) batteries.

At the current stage, these modules are implemented using equation-based formulations that emphasize physical interpretability and computational efficiency. The design ensures that these DER components can be seamlessly integrated with other systems, such as buildings and HVAC systems, within the BESTOpt simulation loop. Here, we briefly introduce some representative components:

- **PV Panels**

The PV module estimates real-time power output based on key influencing factors such as solar irradiance, temperature derating, soiling (the accumulation of dust, dirt, pollen, bird droppings, and other debris on solar panels, which reduces their energy output by blocking or scattering sunlight), shading, inverter efficiency, and long-term degradation. While simplified, this formulation captures essential dynamics and allows for accurate system-level coordination and grid interaction.

- **Battery Energy Storage Systems**

The battery model tracks the state-of-charge (SOC) over time and accounts for factors such as charging/discharging efficiency, temperature effects, and degradation behavior (due to state of health). This supports accurate evaluation of charge/discharge cycles and long-term energy availability.

- **EVs**

EVs are modeled as mobile energy storage units with SOC dynamics similar to stationary batteries. However, their availability is time-varying, and they can be configured with customizable schedules to simulate charging behaviors, mobility constraints, and vehicle-to-grid (V2G) or vehicle-to-building (V2B) scenarios.

### 2.2.3 Introduction of the Controller Modules

As discussed earlier, BESTOpt adopts a hierarchical control structure that supports both multi-scale (from cluster to component) and multi-domain (e.g., thermal, electric, water) coordination. This design supports a wide range of control strategies, which from simple rule-based logic to model-based control (e.g., MPC) and learning-based approaches (e.g., Reinforcement Learning).

Users have the flexibility to implement and test controllers at any hierarchical level depending on their application needs. For example, a user may design a high-level controller at the domain or system level, or define local component-level control using a plug-and-play manner. This flexibility is critical for integrating heterogeneous control strategies across multiple systems and applications.

Below, we define the typical responsibilities and design considerations for each controller level:

- **Cluster-Level Controller**

This controller operates at the highest level, with access to global observations across the entire cluster. It is capable of making high-level, cross-domain decisions, such as coordinating HVAC operations with DER scheduling or managing load shifting across buildings. Due to the large observation space and potential interdependencies, designing a cluster-level controller typically involves greater complexity and computation.

- **Domain-Level Controller**

A domain-level controller manages control tasks within a specific energy domain (e.g., thermal, electric, or water). It is particularly useful when multiple systems or buildings are registered under the same domain, enabling centralized optimization within that scope. For instance, a thermal domain controller may coordinate all HVAC units across buildings to optimize energy efficiency or peak load reduction.

- **System-Level Controller**

This controller manages the behavior of an individual system, such as an HVAC system or DER (e.g., battery, PV). This level is the most commonly used for decentralized control, and its output typically serves as a supervisory reference signal for downstream component-level controllers.

- **Component-Level Controller**

At the most granular level, component-level controllers directly operate individual devices such as fans, coils, pumps, PVs, and batteries. These controllers usually track reference signals provided by system-level controllers and apply basic control logic (e.g., PID or rule-based) to actuate the hardware or simulation model accordingly.

To illustrate this hierarchical coordination, we use a fan-coil-unit (FCU) based HVAC system to demonstrate the interaction between the system-level and component-level controllers, as shown in Figure 5. In the FCU system, two component-level controllers have been implemented: the fan controller for regulating airflow rate and the pump controller for regulating water flow rate. Both support linear and staged control logics.

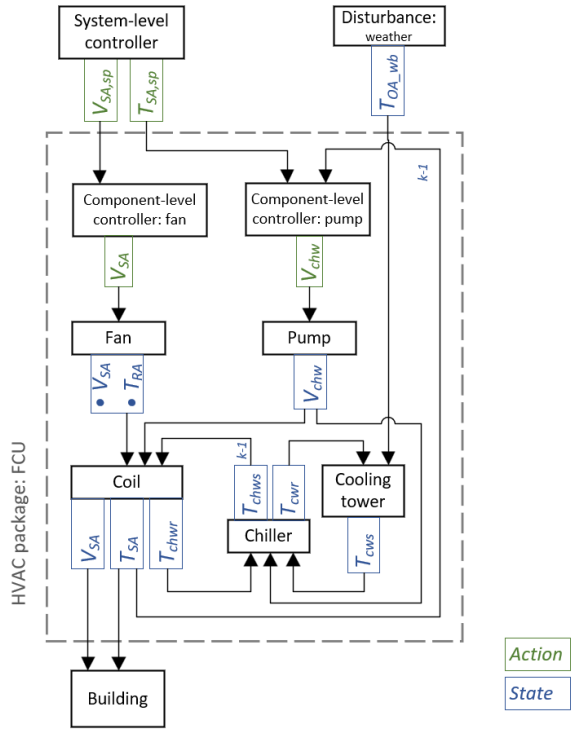

*Figure 5 Example of hierarchy HVAC control, where V for flow rate and T for temperature. SA for supply air, RA for return air, chw for chilled water, chws for chilled water supply (from the chiller), and chwr for chilled water return (to the chiller), cws for cooling water supply, cwr for cooling water return, OA_wb for outdoor air wet bulb temperature.*

### 2.2.4 Introduction of the Disturbance Modules

The disturbance module in BESTOpt simulates exogenous inputs that update over the simulation time steps and affect both system behavior and control decisions but are not directly controlled by the user. The framework currently supports three key categories of disturbances:

- **Occupancy**

Occupancy is modeled across three dimensions to capture its diverse influence on building operations:
**Mobility** reflects occupancy status at the building level (e.g., occupied vs. unoccupied), which affects things such as HVAC usage and EV availability.
**Comfort** relates to occupants' thermal expectations and preferences, which influence thermal setpoints and lighting demand.
**Activity** captures occupants' interaction with appliances and equipment, directly impacting plug loads, domestic hot water use, and internal heat gains.

- **Weather**

The weather module provides environmental inputs such as outdoor air temperature and solar radiation, which directly impact building thermal dynamics, PV generation, and battery operation.

- **Price Signals**

The price signal module supports multiple pricing schemes depending on the case study. This includes dynamic pricing (e.g., real-time electricity market prices) and fixed patterns such as Time-of-Use (TOU) pricing. These signals are used to inform cost-aware control strategies, support demand response coordination, and evaluate economic performance under various tariff structures.

These modules can be configured in multiple ways:
1) by loading predefined profiles (e.g., from historical ".CSV", ".EPW" datasets),
2) by loading form online resources (e.g., weather API),
3) by dynamically forecasting future values based on historical data.

For forecasting-based disturbance generation, BESTOpt integrates a CNN-LSTM-Bayesian neural network that learns temporal patterns and supports real-time forecasting of disturbance variables. For example, Figure 6 shows an application of this approach for one-day-ahead occupancy number forecasting, but the same architecture can be applied to other variables such as weather or price signals (Jiang, Z., & Dong, B. 2025). This framework supports flexible forecasting across multiple disturbance types, offering a unified method for predictive based simulation and control optimization.

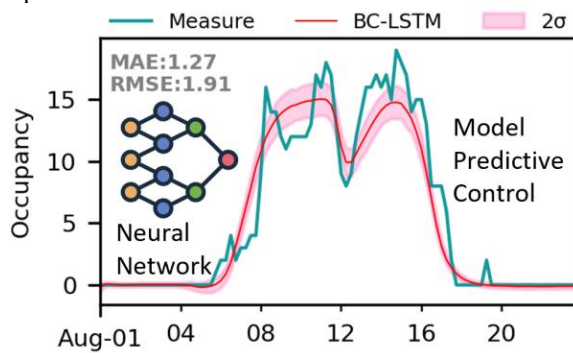

*Figure 6 CNN-LSTM-Bayesian Neural Network for one day ahead occupancy forecasting*

Additionally, we provide an optional integration with our open-source RESI package (Jiang, Z., & Dong, B. 2024), which enables the generation of future weather data under typical scenarios and extreme conditions to support resilience-focused studies.

This flexible and extensible disturbance module design ensures that BESTOpt can support a wide range of simulation and control tasks from standard operations to uncertainty-aware planning under extreme conditions.

## 3 Case Study

### 3.1 Component Level Control of a Single-zone Building with FCU Systems

Figure 7 compares the differences between control actions and actual system feedback using a single-zone building equipped with an FCU system as an example. Due to configuration issues, mismatches between the controller's commands and real system responses are common in real-world building settings. In this example, we evaluated three types of fans (constant-speed, staged, and variable frequency drive (VFD)) to test their ability to track the control signal. The constant-speed fan only operates in on/off modes and shows noticeable deviation from the action signal; similarly, the staged fan cannot perfectly match the control input, while the VFD fan can closely follow it. This example highlights how BESTOpt captures the true system dynamics, unlike many existing platforms which rely on predefined curves and neglect real-world dynamics, often leading to unrealistic results.

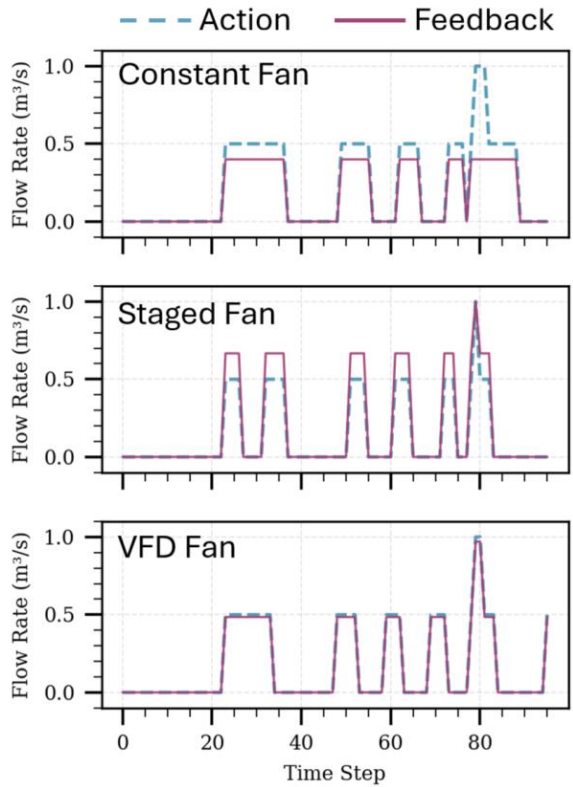

*Figure 7 Comparison of control actions and real feedback across different fan types*

### 3.2 Building Thermal Dynamic Performance

The predictive accuracy (Jiang, Z., & Dong, B. 2024, 2025) and control performance (Jiang, Z., & Dong, B. 2025) of BESTOpt's building thermal dynamic models have been thoroughly evaluated in our previous studies. Here, we present a simple case study to highlight the significance of embedding physics priors into machine learning models, as illustrated in Figure 8.

In this example, the green line represents the ground truth zone air temperature, while the red lines show the predicted temperature over a 24-hour horizon with 15-minute intervals. Both the LSTM baseline and the PI-ModNN model demonstrate good performance on July 30th, when the HVAC system is operating under normal conditions.

However, to evaluate model generalization under unseen scenarios, the HVAC system is intentionally turned off after July 30th. As a result, indoor temperature continues to rise due to ambient heat gains during the summer. In this scenario, the LSTM baseline fails to capture the continued thermal drift, significantly underestimating the zone temperature. In contrast, the PI-ModNN model in BESTOpt accurately reflects the ongoing thermal dynamics, maintaining physical consistency and yielding much lower prediction errors. This result demonstrates the critical value of incorporating physics-informed constraints into data-driven models, especially for control-oriented applications, where the accuracy of control response is essential, and also for RL agent training, where the agent must explore model responses under unseen or abnormal conditions.

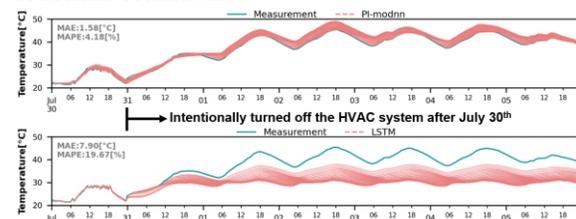

*Figure 8 model performance in unseen scenarios*

### 3.3 Model and Control of a Single-Family House with Fan Coil Unit and PV-Battery system

Figure 9 presents a one-day example from a case study involving a single-family house (based on an EnergyPlus prototype) equipped with a FCU for HVAC and a DER system consisting of a 5 kW PV array, a 10 kWh battery, and two EVs with 60 kWh and 40 kWh capacities (sizes only used for demonstration). This case demonstrates the integrated operation of building dynamics, HVAC control, and DER coordination under rule-based strategies.

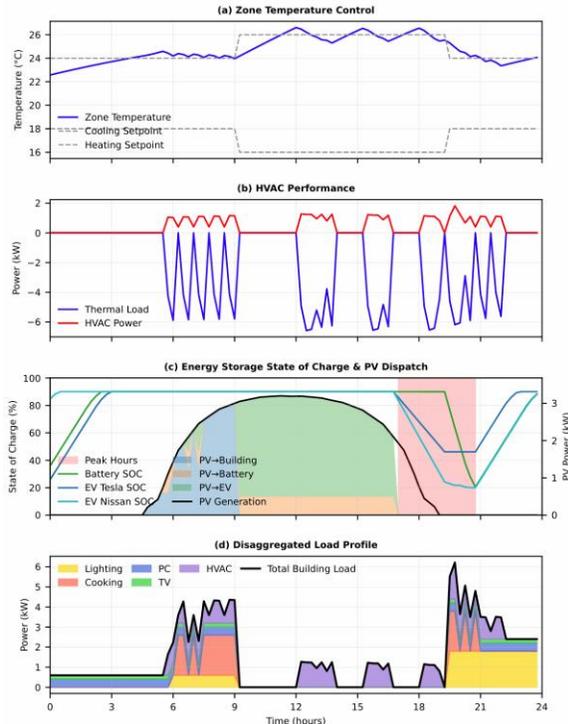

*Figure 9 One day simulation example of a single-family house with fan-coil unit HVAC system and PV-Battery-EV DER system*

- Subplot A illustrates the building thermal dynamics, where the zone air temperature is controlled using a rule-based ON–OFF controller with a deadband. This controller maintains thermal comfort by adjusting the HVAC operation based on indoor temperature relative to the setpoints.
- Subplot B shows the HVAC system's thermal load and total power, including the fan, pump, coil, and other mechanical subsystems. This disaggregation highlights the operational behavior of each component and its contribution to overall energy demand.
- Subplot C depicts the DER operation, where the top graph tracks the state of charge (SOC) of both the battery and EV. A simple rule-based control strategy is applied, driven by a Time-of-Use (TOU) price signal: charging is prioritized during off-peak hours, and discharging occurs during peak hours (red shaded area). The bottom shaded area shows the total PV generation, along with a rule-based dispatch strategy: PV power is first allocated to meet building demand, then to charge the EV, and finally to charge the stationary battery.
- Subplot D presents the disaggregated building load profile, capturing not only HVAC-related demand but also lighting and occupant-driven activities such as cooking, TV use, and computer usage. This breakdown demonstrates the import role of human behavior on energy consumption.

This example demonstrates BESTOpt's capability to simulate fine-grained interactions between buildings, systems, and DERs, and to evaluate control logic under realistic operation scenarios.

### 3.4 Model and Control of 5 Single-Family Houses with Fan Coil Units

Figure 10 presents a two-day simulation example of a small residential cluster composed of five single-family houses, each equipped with a FCU. This case demonstrates BESTOpt's capacity to model and coordinate cluster-level control and analysis.

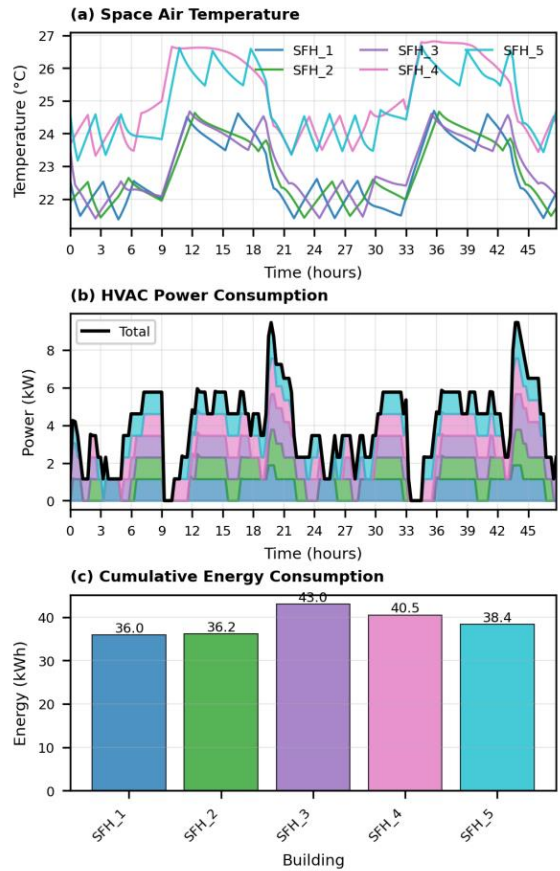

*Figure 10 Two-day simulation example of a five-building residential cluster, each with FCU HVAC systems, highlighting model scalability, load diversity, and cross-building comparisons*

- Subplot A shows the thermal dynamics of five buildings, each modeled using PI-ModNN trained on just three months of historical data, with model training times of under five minutes per building. Despite their differences, all zones are successfully maintained within their individual setpoint ranges, highlighting the generalizability and adaptability of the modeling approach.
- Subplot B presents the total HVAC power consumption for the cluster, along with the individual load profiles of each FCU system. This disaggregated view enables clear comparisons across buildings and supports the design of advanced controllers at different hierarchical levels.
- Subplot C shows the cumulative energy consumption of the cluster over two days. Differences across the five houses emerge due to variations in envelope characteristics, building size, and HVAC configuration (e.g., setpoints). These insights demonstrate BESTOpt's potential for use cases such as energy benchmarking, retrofit evaluation, and scenario-based decision-making.

## 4. Discussions

This study presents the structure and capabilities of BESTOpt, a modular, physics-informed machine learning-based runtime environment designed to support multi-scale and multi-domain coordination across occupancy, building dynamics, HVAC systems, and distributed energy resources. By adopting a hierarchical module structure and a standardized state–action–disturbance–observation data typology, BESTOpt enables integrated modeling, control, and optimization for complex building energy systems. Several key points warrant further discussion:

**1) Trade-off Between Physics-Based and Data-Driven Models**
Our results demonstrate the advantage of embedding physics priors into machine learning for building thermal dynamics, particularly in terms of generalization ability under unseen conditions.
However, the benefits of PIML are less clear for HVAC and DER components, where traditional first-principles models or purely data-driven models may already offer sufficient accuracy, depending on the application.
Important open questions include:

- When is a pure physics-based model sufficient?
- Under what conditions do data-driven models outperform?
- What level of physical constraint should be embedded in a hybrid model, and at which application scenarios?

These questions require further evaluation and domain-specific benchmarks to guide the future development of PIML-based modules.

**2) Modularity and Extensibility**
BESTOpt is designed to be an extensible platform. The use of standardized base classes and structured input/output interfaces enables a wide range of users to collaborate by contributing their own models. Different types of modules such as dynamic, control, or disturbance can be tested, replaced, updated and compared within the unified framework.

**3) Current Limitations and Future Directions**
While the current thermal dynamic model supports both single-zone and multi-zone buildings (up to 30 zones), we observe that training time increases significantly in multi-zone cases. This scalability challenge represents a potential limitation and a future optimization target. Additionally, for high-level controllers (e.g., domain or cluster level), the observation and action spaces become high-dimensional, making model training and optimization more complex. Advanced dimensionality reduction techniques, hierarchical policy structures, or multi-agent coordination strategies may help address this challenge.

## 5. Conclusions

This paper presents BESTOpt, a modular, physics-informed machine learning framework designed to address critical challenges in future building benchmarking and evaluation, sensing, control and optimization, building performance simulation, and measurement and verification. By combining a structured hierarchical architecture (cluster–domain–system/building–component) with a standardized data typology (state–action–disturbance–observation), BESTOpt enables scalable integration of occupancy, building dynamics, HVAC systems, DERs, and grid-level interactions.

Our results highlight the practical value of embedding physics priors into machine learning models—particularly for building thermal dynamics under unseen scenarios where generalization and physical consistency are critical. The framework's modular architecture supports plug-and-play development of dynamic, control, and disturbance modules, allowing researchers to rapidly prototype, compare, and benchmark diverse modeling and control strategies. Case studies further demonstrate its flexibility across both single-building and cluster-scale applications, and its support for both centralized and decentralized control schemes.

Future work will focus on expanding BESTOpt's capabilities for efficient multi-zone training, advanced controller development, enhancing controller scalability under high-dimensional observation and action spaces, and developing benchmarking protocols for large scale evaluations. We envision BESTOpt as a foundational platform to accelerate innovation in smart, resilient, and decarbonized building ecosystems.

## Acknowledgment

## References

Wetter, M., & Sulzer, M. (2024). A call to action for building energy system modelling in the age of decarbonization. *Journal of Building Performance Simulation*, *17*(3), 383-393.

Jiang, Z., & Dong, B. (2024). Modularized neural network incorporating physical priors for future building energy modeling. *Patterns*, *5*(8).

Jiang, Z., Wang, X., Li, H., Hong, T., You, F., Drgoňa, J., ... & Dong, B. (2025). Physics-informed machine learning for building performance simulation-A review of a nascent field. *Advances in Applied Energy*, 100223.

Jiang, Z., & Dong, B. (2025). *EasyMPC: An open-source toolkit for model predictive control in smart buildings*. 2025 ASHRAE Annual Conference, Phoenix, AZ.

Jiang, Z., & Dong, B. (2024). RESI: A Power Outage Event and Typical Weather File Generator For Future RESIlient Building Design and Operation.

Jiang, Z., Wang, X., & Dong, B. (2025). Physics-informed modularized neural network for advanced building control by deep reinforcement learning. *Advances in Applied Energy, 19,* 100237.

Y. Zhang, M. Liu, Z. Yang, C. Calfa, Z. O'Neill. "Development and Validation of a Water-to-Air Heat Pump Model Using Modelica." The American Modelica 2024 Conference. Storrs, CT, USA. October14-16. 2024.